\documentstyle[sprocl,epsf]{article}

\def\Journal#1#2#3#4{{#1} {\bf #2}, #3 (#4)}

\def\NPB{{\it Nucl. Phys.} B}
\def\PLB{{\it Phys. Lett.}  B}
\def\PRL{\it Phys. Rev. Lett.}
\def\PRD{{\it Phys. Rev.} D}

\def\be{\begin{equation}}
\def\ee{\end{equation}}
\def\bea{\begin{eqnarray}}
\def\eea{\end{eqnarray}}

\begin{document}

\title{TEST OF CPT AND LORENTZ INVARIANCE FROM MUONIUM SPECTROSCOPY}

\author{D. KAWALL, V. W. HUGHES, 
M. GROSSE PERDEKAMP\footnote{Current Address : Riken-BNL Research Center, 
Upton, NY 11973, USA}, W. LIU}

\address{Department of Physics, Yale University,\\
New Haven, CT 06520-8120, USA\\E-mail: david.kawall@yale.edu} 

\author{K. JUNGMANN\footnote{Current Address : KVI, Zernikelaan 25, NI-9747 AA 
Groningen, The Netherlands}, G. ZU PUTLITZ}

\address{Physikalisches Institut, Universit\"at Heidelberg,\\ 
D-69120 Heidelberg, Germany}


\maketitle\abstracts{ 
Following a suggestion of Kosteleck\a'{y} {\it{et al.}}
we have evaluated a test of CPT and Lorentz invariance from the microwave
spectroscopy of muonium. 
Violations of CPT and Lorentz invariance 
in the muon sector would be indicated by
frequency shifts $\delta\nu_{12}$ and $\delta\nu_{34}$ in
$\nu_{12}$ and $\nu_{34}$, 
the two hyperfine transitions involving muon spin flip, which were 
precisely measured in ground state muonium in a strong magnetic field 
of 1.7 T. Such shifts would appear in the laboratory frame as
anti-correlated oscillations in $\nu_{12}$ and $\nu_{34}$ at
the earth's sidereal frequency. Our experiment found no time dependence 
in $\nu_{12}$ or $\nu_{34}$ at the level of 20 Hz, limiting 
the size of some CPT and Lorentz violating parameters 
at the level of $2\times10^{-23}$ GeV, representing for the moment the
most sensitive limits on some of the muon parameters of the theory.
}

\section{Introduction}
Muonium ($\mu^{+}e^{-},~M$) is the hydrogenlike bound state of a positive muon
and electron. Interest in this simple atom arises for several reasons.
For precision tests of QED, muonium is well suited since for instance,
unlike for hydrogen, its ground state hyperfine structure can be calculated
with high precision {\hbox{since}} the complications of hadronic 
structure are absent. The combination of precise calculations 
and experiments enable precision tests of QED to be performed, and 
also allows the accurate determination of 
fundamental constants \cite{muonium,m1s2s,mohr}. Muonium is also used
in searches for physics beyond the standard model, specifically lepton number
violation \cite{mmbar}, since it combines leptons from two different
generations.

Another test of fundamental physics became possible after the development
by Kosteleck\a'{y} and coworkers \cite{kos1,kos2,kos3} of
a plausible mechanism by which CPT and Lorentz symmetry might be
violated, based on spontaneous breaking of CPT and Lorentz symmetry in an 
underlying higher order theory. The new phenomenology resulting from CPT and 
Lorentz violation was incorporated in extensions to the standard model, 
and its implications for muonium, which prompted this work
(more details of which may be found in \cite{muoniumcpt}), were described 
in \cite{bkl} and R. Bluhm's talk at this conference. 

\begin{figure}
\centering
\epsfxsize=\linewidth
\epsfbox{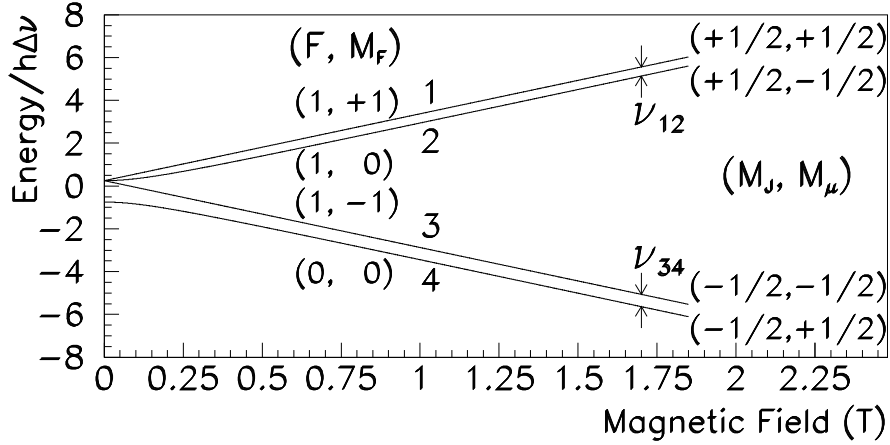}
\caption{Breit-Rabi energy level diagram of ground state muonium. At
high fields, the indicated transitions, $\nu_{12}$ and $\nu_{34}$ are 
essentially muon spin flip transitions.}
\label{fig:mbreitrabi}
\end{figure}

The QED extension of the theory leads to potentially
observable perturbations in the $1^{2}S_{1/2}$ ground state energy levels of 
muonium, where in a strong field, the ground state splits into four substates 
labelled 1 through 4 in order of decreasing energy, defined
by the magnetic quantum numbers $(M_J,M_{\mu})$ (see Fig.\ref{fig:mbreitrabi}).
Of the six possible
ground state hyperfine transitions we focus on 
$({\textstyle{1/2,1/2}})$ $\leftrightarrow$ $({\textstyle{1/2,-1/2}})$
designated $\nu_{12}$,
and $({\textstyle{-1/2,-1/2}})$ $\leftrightarrow$ $({\textstyle{-1/2,1/2}})$ 
designated $\nu_{34}$, which were
are observed by a microwave magnetic resonance technique \cite{muonium}. The 
transition frequencies are given by the Breit-Rabi formula
\begin{eqnarray}
\nu_{12}=-\frac{\mu_{B}^{\mu}g'_{\mu}{H}}{h}+\frac{\Delta\nu}{2}
\left[\left(1+x\right)-\sqrt{1+x^2}\right],\label{eqn:nu12}\\
\nu_{34}=+\frac{\mu_{B}^{\mu}g'_{\mu}{H}}{h}+\frac{\Delta\nu}{2}
\left[\left(1-x\right)+\sqrt{1+x^2}\right],\label{eqn:nu34}
\end{eqnarray}
where $x=\left(g_{J}\mu_{B}^{e}+g'_{\mu}\mu_{B}^{\mu}\right)H/(h\Delta\nu)$ 
is a dimensionless parameter proportional to the magnetic field strength, $H$,
and $\Delta\nu$ is the ground state hyperfine interval.
We used the Larmor relation, $2\mu_{p}H=h\nu_{p}$, and NMR to determine $H$ 
in terms of the free proton precession frequency, $\nu_{p}$, and the proton 
magnetic moment, $\mu_{p}$. The electron and muon $g$ factors in muonium, 
$g_{J}$ and $g'_{\mu}$ respectively, differ from their free space by
binding corrections of order $\alpha^{2}$ (see \cite{muonium}).
The experiment, performed at the Clinton P. 
Anderson Meson Physics Facility at Los Alamos,  
measured the ground state hyperfine interval 
$\Delta\nu=\nu_{12}+\nu_{34}$ to 12 ppb precision
to compare with theory as a precise test of
QED, and also, using (\ref{eqn:nu12}), (\ref{eqn:nu34})  
and the measurements of the transition frequencies $\nu_{12}$ and $\nu_{34}$
(each made to about 40 Hz uncertainty ($\approx$ 20 ppb)) 
to extract a value of $\mu_{\mu}/\mu_{p}$ (to $\approx$ 120 ppb) for 
positive muons.

At the strong field of 1.7 T used in the experiment, 
$x~\simeq~10.7~\gg~1$, these transitions correspond essentially to pure muon 
spin flip. Lorentz violating energy shifts, 
$\delta\nu_{12}$ and $\delta\nu_{34}$, 
to the transition frequencies, $\nu_{12}$ and $\nu_{34}$, in muonium coming 
from the standard model extension, may then be attributed to the muon
parameters alone of the theory. The prediction, described in detail 
in \cite{bkl} is :
\begin{eqnarray}
\delta\nu_{12}~\approx~-\delta\nu_{34}~
\approx~\tilde{b}_{3}^{\mu}/\pi,
\label{eqn:prediction}
\end{eqnarray}
where $\tilde{b}_{3}^{\mu}~\equiv~b_{3}^{\mu}~+~d_{30}^{\mu}m_{\mu}~+~
H_{12}^{\mu}$ are laboratory frame parameters. 
Precision microwave spectroscopy on muonium can measure or set limits on 
these symmetry violating terms, with sensitivity at
the Planck scale level~\cite{bkl}. 

Predicting $\nu_{12}$ and $\nu_{34}$ from QED requires knowledge of many atomic
constants; $m_{\mu}$, $\mu_{\mu}$, and $\Delta\nu$ in particular, as well as 
the calculation of higher order QED radiative corrections,
and small electroweak and hadronic radiative corrections. The relevant 
constants (and hence calculation results) are not known to as high accuracy as 
the experimental determinations of $\nu_{12}$ and
$\nu_{34}$. Comparing
predictions for $\nu_{12}$ and $\nu_{34}$ (based on independent determinations
of the required atomic constants) with the experimental results has
poor sensitivity to the non-standard model energy shifts
$\delta\nu_{12}$ and $\delta\nu_{34}$. 

The most powerful signature of CPT and Lorentz violation in this case
comes from the observation that since the laboratory
rotates with the earth, and the parameters of CPT and Lorentz violation
involve spatial components in a fixed celestial frame, the experimentally 
observed $\nu_{12}$ and $\nu_{34}$ may oscillate about a mean value
at the earth's sidereal frequency 
$\Omega~=~2\pi/23~{\mathrm{hr}}~56~{\mathrm{min}}$ with amplitudes 
$\delta\nu_{12}$ and $\delta\nu_{34}$. An experimental limit on the oscillation
amplitude $\delta\nu_{12}$ implies constraints on the celestial frame
parameters :
\begin{eqnarray}
{\frac{1}{\pi}}|{\mathrm{sin}}\chi|
\sqrt{ \big({\tilde{b}_{X}^{\mu}}\big)^2+
       \big({\tilde{b}_{Y}^{\mu}}\big)^2} &\leq& \delta\nu_{12}
\label{eqn:constraint}
\end{eqnarray} 
in which $\chi\sim~90^{\circ}$ is the angle between the earth's
rotational axis, $\hat Z$, and the 
quantization axis defined by the laboratory magnetic field at 
Los Alamos (parameters and coordinates are described in \cite{bkl}).

The experimental signature has several nice features.
The sum of the transition frequencies, $\nu_{12}+\nu_{34}$, is
equal to the ground state hyperfine splitting, $\Delta\nu$, and since we 
expect (see Eqn.\ref{eqn:prediction})
$\delta\nu_{12}+\delta\nu_{34}\approx~0$, no sidereal variation is
expected in the hyperfine interval. However, the 
difference in transition frequencies
$\delta\nu_{12}-\delta\nu_{34}\approx~2\tilde{b}_{3}^{\mu}/\pi$
would exhibit the maximum sidereal variation. The expectation that sidereal
variations in $\nu_{12}$ and $\nu_{34}$ would be exactly out of phase allows 
the above consistency check against spurious signals. We also note that
at strong fields, $\nu_{34}-\nu_{12}$ is almost proportional to the
magnetic moment of the muon, so we are essentially probing for a 
sidereal variation in the magnetic moment (while the $g$ factor stays
constant to first order in the theory).

\section{Details of the Experiment and Analysis} 
\begin{figure}
\centering
\epsfysize=7cm
\epsfbox{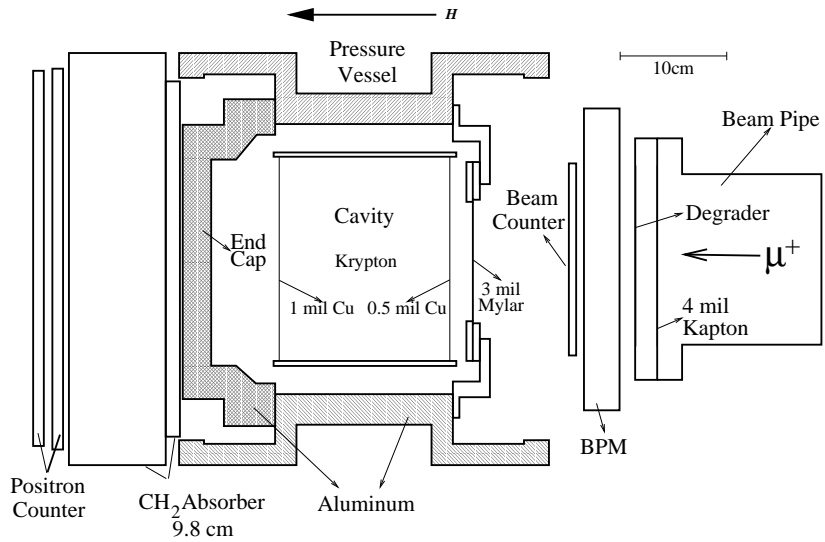}
\caption{Schematic of the experimental apparatus.}
\label{fig:apparatus}
\end{figure}

\begin{figure}
\centering
\epsfysize=7cm
\epsfbox{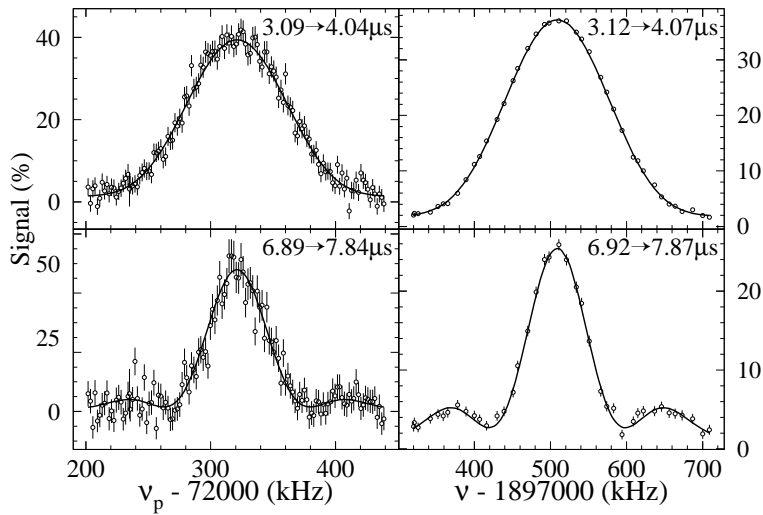}
\caption{Muonium resonance lines (data and fit) for $\nu_{12}$ taken using
magnetic field sweep on the left, and microwave frequency sweep on 
the right, for muonium atoms which have decayed in selected time
intervals after formation.}
\label{fig:reslines}
\end{figure}

The measurements of $\nu_{12}$ and $\nu_{34}$ were done in
a microwave magnetic resonance experiment \cite{muonium}. 
Muonium was formed by electron capture by polarized muons (negative helicity)
stopping in a 
krypton gas target. The muons would subsequently decay weakly via
$\mu^{+}\rightarrow e^{+}\nu_{e}\bar{\nu}_{\mu}$ where the momentum and
angle of the $e^{+}$ is a function of the muon polarization. Since high momenta
decay positrons are emitted preferentially in the muon spin direction,
by driving the muon spin flip transitions $\nu_{12}$ and $\nu_{34}$ with
a microwave magnetic field perpendicular to the strong static field,
the angular distribution of high momenta $e^{+}$ could be changed from
predominantly upstream to downstream with respect to the beam direction
if the microwave field was near the resonance frequency for the transition
for the given value of the static field. The apparatus is shown in 
Fig. \ref{fig:apparatus}.
Resonance lines were observed by varying the magnetic field with fixed 
microwave frequency and by varying the microwave frequency with fixed 
magnetic field. 
A line narrowing technique \cite{boshier} was used involving observation
of a transition signal only from $M$ atoms which have lived
considerably longer than $\tau_{\mu}~\sim~2.2~\mu$s  (Fig. \ref{fig:reslines}).

To search for a time dependence of $\nu_{12}$ and $\nu_{34}$,
data from each resonance line run 
(each lasting about half an hour) were fit at the 
measured magnetic field strength and Kr pressure to determine 
provisional line centers for $\nu_{12}$ and $\nu_{34}$. These line
centers were transformed to their values in a magnetic field strength
corresponding to a free proton precession frequency of 72.320 000 MHz.
The data were taken at Kr pressures of 0.8 and 1.5 atm, so the line centers
were corrected for a small quadratic pressure shift, and were 
extrapolated linearly to their values at zero pressure, using a 
pressure shift coefficient determined from the data. 

The overall results were 
$\nu_{12}({\mathrm{exp}})~=~1~897~539~800~(35)~{\mathrm{Hz}}~
(18~{\mathrm{ppb}})$ and $\nu_{34}({\mathrm{exp}})~=~2~565~762~965~(43)~
{\mathrm{Hz}}~(17~{\mathrm{ppb}})$ where the uncertainties reflect 
statistical and systematic uncertainties combined. 
Some systematic
uncertainties which were pertinent for the extraction of these final
vacuum values of $\nu_{12}$ and $\nu_{34}$ for tests of 
QED and for extracting constants, such as uncertainties in
the absolute calibration of the pressure meters or pressure shift
coefficients, do not affect the extraction of possible sidereal 
variations at first order.
The results for $\nu_{12}$ and $\nu_{34}$ were then grouped
as a function of sidereal time, where time zero has been
set as the time in 1995 when we obtained our first data. 
\begin{figure}
\centering
\epsfysize=7cm
\epsfbox{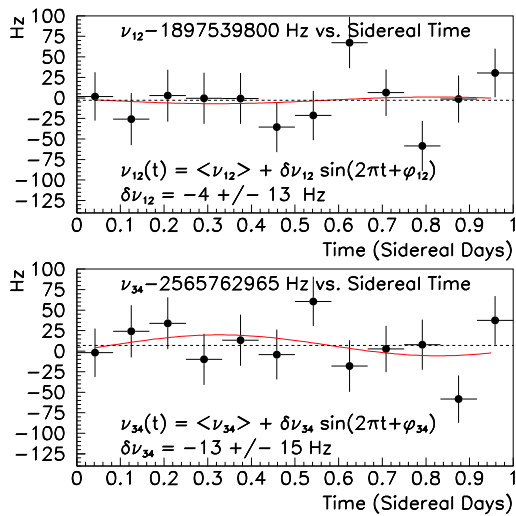}
\caption{Two years of data on $\nu_{12}$ and $\nu_{34}$ are shown binned 
versus sidereal time and fit for a possible sinusoidal variation. The
amplitudes are consistent with zero.}
\label{fig:muonium_nuij}
\end{figure}
The data obtained in 1995 and 1996 are plotted as a function of 
time measured as a function of a sidereal day in Fig. \ref{fig:muonium_nuij},
where twelve points at $\approx$ 2 hr. intervals are plotted, and the 
vertical scale is in Hz. The data for $\nu_{12}$ and $\nu_{34}$ 
were fit separately by the functions
\begin{eqnarray}
\nu_{ij}(t)=\langle\nu_{ij}\rangle+\delta\nu_{ij}\sin
\left(2\pi t+\phi_{ij}\right)
\label{eqn:funcform}
\end{eqnarray}
where $t$ is the time in sidereal days, and the 
fit parameters are $\langle\nu_{ij}\rangle$, 
the amplitude of the possible time 
variation $\delta\nu_{ij}$, and the
phase $\phi_{ij}$ (where no phase relation was assumed between 
$\nu_{12}$ and $\nu_{34}$).

\section{Potential Systematic Effects}
Non-zero values for $\delta\nu_{12}$ and $\delta\nu_{34}$ could have
arisen from systematic effects which lead to 
variations in the parameters determining the line centers - especially
parameters varying with a period of $\approx$ 24 hr. Principal concerns are
possible day-night variations of the magnetic field strength, and
of the density and temperature of the Kr gas stopping target.

Diurnal variations of several $^\circ$C
in the experimental hall lead to oscillations in the
magnetic field strength of the persistent-mode superconducting solenoid of
$\leq$ 1 ppm. Changes of 0.05 ppm 
in the field strength were easily resolved, and the oscillation's  
effects on the line centers were accounted for in extracting the
line centers. Temperature
changes also affected the diamagnetic shielding constant of the water in the
NMR probes used to monitor the field
(the probes were not temperature-stabilized, but were in good thermal 
contact with 
the microwave cavity which was temperature-stabilized to 0.1~$^\circ$C). 
A conservative upper limit of 2~$^\circ$C diurnal
variation in water temperature (estimated from a simple thermal transport
model) would change the NMR frequencies by 0.02 ppm, leading to errors in the
line centers of about 2.5 Hz; of opposite sign for $\nu_{12}$ and
$\nu_{34}$. This potential effect is well below the statistical sensitivity
for sidereal variations of 12 to 15 Hz. 

The effect of the diurnal variation of Kr pressure has been
evaluated. The 76 $\mu$m mylar
front end window to the Kr stopping target flexed with day-night 
variations of the external atmospheric pressure.
This induced fractional oscillations in the Kr gas target 
pressure which were measured to be 
about $2.5 \times 10^{-4}$.
Through pressure shift coefficients
of about -16.5 kHz/atm for $\nu_{12}$ and -19.5 kHz/atm for $\nu_{34}$, 
the resulting shifts in the line centers (typically 
7.5 Hz in $\nu_{34}$ and 6 Hz in $\nu_{12}$) were automatically accounted 
for in performing the extrapolation to zero pressure, and do not
contribute any significant time variation to $\nu_{ij}$.

The pressure shift coefficients depend on the average velocities of the 
{\hbox{atoms}}, and so are functions of
temperature. The fractional changes in the transition frequencies with
temperature (measured in hydrogen and its isotopes~\cite{morgan}) 
are roughly 
$1\times 10^{-11}~^{\circ}{\mathrm{C}}^{-1}{\mathrm{Torr}}^{-1}$. Given the
temperature stability of the Kr gas of about 0.1~$^\circ$C, 
temperature dependent errors introduced into the extrapolation of the 
line centers to their vacuum values would be limited to a few Hz, below the 
statistical sensitivity of our test.

Other potential concerns involve the two frequency references used
in the experiment - the proton precession frequency forming the
basis of the magnetic field determination, and the Loran-C 10 MHz
frequency reference used for the NMR and microwave frequency synthesizers. 
The Loran-C standard is based on hyperfine transitions in Cs with $m_F$=0,
and so is insensitive to any preferred spatial orientation, and would 
not introduce a signature for Lorentz violation into the spectroscopic
measurements. Significant oscillations in the Loran-C are precluded by
the null results in both $\nu_{12}+\nu_{34}$ and $\nu_{34}-\nu_{12}$.
Finally, bounds on clock comparisons of 
$^{199}$Hg and $^{133}$Cs \cite{clock,berglund} 
place limits on the Lorentz violating energy shifts in the precession
frequency of a proton of $10^{-27}$ GeV, which
imply the NMR measurements are free of shifts well below the Hz level.

\section{Results}
\begin{figure}
\centering
\epsfysize=7cm
\epsfbox{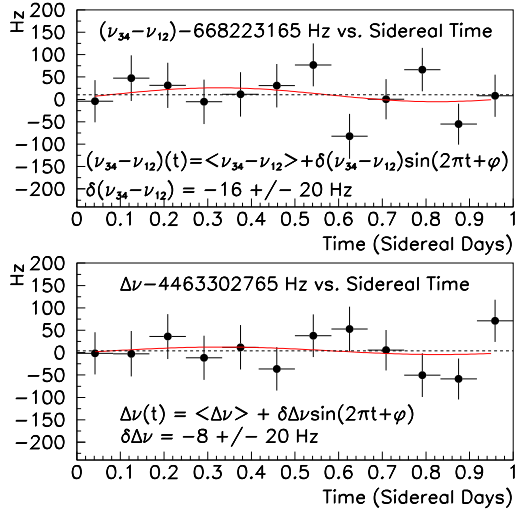}
\caption{Two years of data on $\nu_{12}$-$\nu_{34}$ and
$\nu_{12}+\nu_{34}=\Delta\nu$ are shown binned 
versus sidereal time and fit for a possible sinusoidal variation. The
amplitudes are consistent with zero.}
\label{fig:muonium_diff}
\end{figure}

The amplitudes for $\delta\nu_{12}$ and $\delta\nu_{34}$ independently 
are consistent with zero, $-4\pm 13$ Hz and $-13\pm 15$ Hz
respectively. However, the standard model extension predicts the phase 
relation $\delta\nu_{12}~\approx~-\delta\nu_{34}$. Plots of 
$\nu_{12}~\pm~\nu_{34}$ versus
sidereal time are shown in Figure \ref{fig:muonium_diff}, and fit for
a sinusoidal variation (as in Eqn.\ref{eqn:funcform}) where a common
phase is assumed between $\nu_{12}$ and $\nu_{34}$. No sinusoidal 
variation with sidereal time within $\pm~20~$Hz is found in either the
sum or difference, which yields a
68$\%$ confidence level (one sigma) limit on the non-rotating frame components 
(see Eqn. \ref{eqn:constraint})
\begin{eqnarray}
\sqrt{ \big({\tilde{b}_{X}^{\mu}}\big)^2+
       \big({\tilde{b}_{Y}^{\mu}}\big)^2} &\leq& 
       2\times 10^{-23}~{\mathrm{GeV}}.  
\end{eqnarray} 
The figure of merit of these results as a test of CPT violation is taken as :
\begin{eqnarray}
2\sqrt{ \big({\tilde{b}_{X}^{\mu}}\big)^2+
       \big({\tilde{b}_{Y}^{\mu}}\big)^2 }/{m_{\mu}}
&\leq& 
       \frac{2\pi\mid\delta\nu_{12}\mid}{m_{\mu}}~\approx~5\times 10^{-22}.
\end{eqnarray}
which is comparable to the 
dimensionless scaling factor $m_{\mu}/M_{P}~\sim~10^{-20}$.
The limits on $\delta\nu_{34}$ and $\delta(\nu_{34}-\nu_{12})$ 
yield similar values.   

\section{Conclusions}
No unambiguous violation of CPT or Lorentz invariance has been observed
in this system, or any other, despite many sensitive tests performed
over the more than 40 years since the first experiments of Hughes and 
Drever\cite{hughesdrever}. 
The null results presented here are the first test for sidereal variations
in the interactions of the muon arising from CPT and Lorentz violation, 
and set the first limits on the associated combination of muon parameters 
of the standard model extension at the level of $2~\times~10^{-23}$ GeV,
representing Planck scale sensitivity, and almost an order of magnitude
improvement in sensitivity over previous results in the muon sector~\cite{bkl}.
Further improvement in the muon parameters may come from the ongoing 
muon g-2 experiment (see M. Deile's contribution in this volume). Improvement
from a new muonium experiment measuring ground state hyperfine transitions
will likely require a more intense muon source,
as the current experiment was statistics limited.

\section*{Acknowledgments}
Our thanks go to V.A. Kosteleck\a'{y} for helpful discussions and
for organizing such an interesting meeting. Thanks also to the
U.S. DOE and BMBF (Germany) for supporting this research.


\section*{References}
\begin{thebibliography}{99}
\bibitem{muonium}W. Liu {\it{et al.}}, {\Journal{\PRL}{82}{711}{1999}}
~and~references~therein.
\bibitem{m1s2s}V. Meyer {\it{et al.}}, \Journal{\PRL}{84}{1136}{2000}.
\bibitem{mohr}P.J. Mohr and B.N. Taylor, Rev. Mod. Phys. {\bf{72}}, 351 (2000).
\bibitem{mmbar}L. Willmann {\it{et al.}}, \Journal{\PRL}{82}{49}{1999}.
\bibitem{kos1}
V.A. Kosteleck\a'{y} and R. Potting, \Journal{\NPB}{359}{545}{1991};
\Journal{\PLB}{381}{89}{1996}.  
\bibitem{kos2} 
D. Colladay and V.A. Kosteleck\a'{y}, \Journal{\PRD}{55}{6760}{1997};
\Journal{\PRD}{58}{116002}{1998}.
\bibitem{kos3}V.A. Kosteleck\a'{y}, ed., {\it{CPT and Lorentz Symmetry}}  
(World Scientific, Singapore, 1999).
\bibitem{muoniumcpt}V.W. Hughes {\it{et al.}}, 
\Journal{\PRL}{87}{111804}{2001}.
\bibitem{bkl}R. Bluhm, V.A. Kosteleck\a'{y}, and C.D. Lane,
\Journal{\PRL}{84}{1098}{2000}.
\bibitem{boshier}M.G. Boshier {\it{et al.}}, 
Phys. Rev. A {\bf{52}}, 1948 (1995).
\bibitem{morgan} C.L. Morgan and E.S. Ensberg, Phys. Rev. A {\bf{7}}, 
1494 (1973).
\bibitem{clock} V.A. Kosteleck\a'{y} and C.D. Lane, 
\Journal{\PRD}{60}{116010}{1999}.
\bibitem{berglund} J.D. Prestage {\it{et al.}},
\Journal{\PRL}{54}{2387}{1985},
C.J. Berglund {\it{et al.}},
\Journal{\PRL}{75}{1879}{1995} and references therein.

\bibitem{hughesdrever} V.W. Hughes, H.G. Robinson and V. Beltran-Lopez, 
\Journal{\PRL}{4}{N504}{1960}, 
R.W.P. Drever, Philos. Mag. {\bf{6}}, 683 (1961).
\end{thebibliography}
\end{document}